
\magnification=1200
\baselineskip=20pt
\tolerance=100000
\overfullrule=0pt

\centerline{\bf SYMMETRY UNDER $\alpha \rightarrow \alpha + 1$
IS FORBIDDEN}
\centerline{\bf  BY HELICITY CONSERVATION}

\vskip 1cm

\centerline{by}

\vskip 1cm

\centerline{C. R. Hagen}
\centerline{Department of Physics and Astronomy}
\centerline{University of Rochester}
\centerline{Rochester, NY 14627}

\vskip 1.2cm

\centerline{\bf \underbar{Abstract}}

The question as to whether helicity conservation in spin one-half Aharonov
Bohm scattering is sufficient in itself to determine uniquely the form of
the spinor wave function near the origin is examined.  Although it is found
that a one parameter family of solutions is compatible with this
conservation law, there must nonetheless be singular solutions which break
the symmetry $\alpha \rightarrow \alpha +1$ required for an anyon
interpretation.  Thus the free parameter which occurs does not allow one to
eliminate the singular solutions even though it does in principle mean that
they can be transferred at will between the spin up and spin down
configurations.

\vfill\eject

\noindent {\bf I. Introduction}

\medskip

The study of spin one-half Aharonov Bohm (AB) scattering has led to a
growing realization that spin can play a decisive role in this phenomenon.
Gerbert$^1$ examined this problem from a somewhat mathematical point of
view, namely the method of self-adjoint extensions.  He considered the spin
up case and noted in the one sector which allowed a normalizable but
singular solution that it was possible in principle to have a one parameter
family of solutions.  The value of this parameter prescribes the relative
contributions of the regular and irregular solutions (i.e.,
 $J_{\pm | m + \alpha |}$ respectively).  This problem was subsequently
considered by this author$^2$ who sought to reformulate the problem as a
limit of a fully realizable physical configuration.  It consisted in
distributing the magnetic flux throughout a cylinder of radius $R$ which
was allowed to go to zero at the end of the calculation.  In contrast to
ref. 1 both the spin up and spin down cases were considered, thereby
allowing one to obtain results for polarized beam experiments.

The results of ref. 2 can be summarized without detailed mathematics.  It
was shown under quite general conditions that for a configuration in which
the magnetic moment interaction is attractive (repulsive) the solution
consists entirely of a singular (nonsingular) function.  Not surprisingly,
this result agreed with that of Gerbert provided that a specific value was
chosen for his free parameter.  Similar results were obtained by Alford
\underbar{et} \underbar{al.}$^3$
 for a single spin projection and under somewhat more restrictive
conditions.

There has been considerable reluctance to accept such results without
reservation since they are not reconcilable to the anyon view.  This is
easily seen by a simple argument.  The singular solutions can occur only
when $\alpha s < 0$ where $s$ is twice the spin projection and $\alpha$ is
the flux parameter.  In this case the sign of the phase shift is
precisely reversed$^2$
relative to the spinless result.  Since the condition $\alpha s < 0$ is not
invariant under the anyonic displacement $\alpha \rightarrow \alpha + 1$,
one clearly has a serious clash between these results and the anyon
interpretation.

To bolster the latter approach various things have been proposed.  One of
these$^4$ sought to use a highly singular nongauge potential to keep the
particle wave function away from the $r < R$ region so that its magnetic
moment interaction could not affect the solution.  This, of course, can be
done and is a perfectly comprehensible quantum mechanical result although
one must do the calculation carefully$^5$ if Klein's paradox is to be
avoided.

Another attempt to circumvent the singular solutions issue makes use of an
appeal to helicity conservation.  It is well known that helicity is
conserved for a Dirac particle in a time independent magnetic field.
Indeed, the solution of ref. 2 has been shown$^6$ to be fully compatible
with this principle.  On the other hand it is not unreasonable to ask (as
has recently been done$^7$) whether the converse is true.  Namely, does
helicity conservation uniquely imply the solution of ref. 2?  That work
(i.e., ref. 7) concluded that there is in fact a one parameter family of
solutions (much as in ref. 1 for the case of a single spin component) and
that the symmetry $\alpha \rightarrow \alpha +1$ could thereby be retained.

This question is reexamined in the present work.  Before presenting the
details it may be well to note at the outset that the conclusion of ref. 7
concerning $\alpha \rightarrow \alpha + 1$ is certainly and obviously
incorrect.  If it were true, then each spin component would have exactly
the same scattering amplitude which would, of course, be equal to the
standard spinless AB amplitude.  That clearly implies an absence of spin
rotation during the scattering process, and that helicity cannot be
conserved.

In the following section the spin one-half AB problem is briefly reviewed
and a two parameter family of solutions derived with no assumptions made
concerning helicity conservation.  The subsequent section uses a direct
calculation of a cross section in a hypothetical scattering experiment to
infer the existence of a relation between these two parameters when
helicity conservation is required.  A concluding section summarizes the
results obtained and offers some general comments on the spin one-half AB
problem.

\vfill\eject

\medskip

\noindent {\bf II. The General Scattering Amplitude}

\medskip

One starts with the Dirac equation in two spatial dimensions
$$E \psi = \left[ M \beta + \beta \gamma \cdot \Pi \right] \psi
\eqno(2.1)$$
where $\Pi_i = - i \partial_i - e A_i$ and
$$e A_i = \alpha \epsilon_{ij} r_j /r^2 \quad .$$
Since both spin up and spin down components are to be included, a convenient
choice for the Dirac matrices is given by
$$\eqalign{\beta &= \sigma_3\cr
\beta \gamma_i &= (\sigma_1 , s \sigma_2) \cr}\eqno(2.2)$$
where $s=\pm1$ (for spin up and spin down) and $\sigma_i$ are the usual
Pauli matrices.  When it is desired to display (2.1) in four-dimensional
form, $s$ should be replaced by $\rho_3$ (namely, the third Pauli matrix
which satisfies $[\rho_3, \sigma_i] = 0$).  In order to be able to
interpret in a physical context the plausibility of the results of this
study the second order form of (2.1) is quite useful.  One finds
$$\left( E^2 - M^2 \right) \psi = \left[ \Pi^2 + \alpha
s \sigma_3 {1 \over r}\ \delta (r) \right] \psi$$
or (in cylindrical coordinates)
$$\left[ {1 \over r}\ {\partial \over \partial r} \ r\ {\partial \over
\partial r} +
{1 \over r^2} \left( {\partial \over \partial \phi} + i \alpha \right)^2 +
k^2 - \alpha s \sigma_3 \ {1 \over r}\ \delta (r) \right] \psi = 0$$
where
$$k^2 \equiv E^2 - M^2 \quad .$$

It should be noted that with the choice (2.2) for the matrix $\beta$ it is
readily seen that the physical (or large component) is $\psi_1$.  Thus upon
expanding $\psi_1$ as
$$\psi_1 = \sum^\infty_{-\infty} g_m (r) e^{im \phi}$$
it follows that
$$\left[ {1 \over r}\ {\partial \over \partial r}\ r\ {\partial \over
\partial r} + k^2 - {(m + \alpha)^2 \over r^2} - \alpha s
\ {1 \over r}\ \delta (r) \right]
g_m (r) = 0 \quad . \eqno(2.3)$$
Clearly the delta function term in (2.3) can be interpreted as a potential
which is repulsive (attractive) for $\alpha s$ greater (less) than zero.
Thus the result of ref. 2 which found a strictly regular (irregular)
solution near the origin in the $R \rightarrow 0$ limit in those cases was
quite reasonable.  In the current context, however, one proceeds in the
spirit of refs. 1 and 7.

Since the solutions of (2.3) must be normalizable at the origin, it follows
that the allowable solutions must be $J_{|m + \alpha |} (kr)$ except
possibly, when $|m + \alpha | < 1$.  Restricting attention to this latter
case one has as the most general solution for the angular momentum state $j
= m + {1 \over 2}\ s$
$$\psi_{_j} = e^{-i(N + {1 \over 2}) \phi - is \phi/2}
\left[ A_s J_{s(\beta - {1 \over 2})-{1 \over 2}} + B_s J_{{1 \over 2} -
s(\beta - {1 \over 2})} \right] \eqno(2.4)$$
where
 $\alpha \equiv N + \beta$ with $N$ the largest integer in
$\alpha$.  Since this implies that $| \beta - {1 \over 2}\ | \leq {1 \over
2}$, it follows that $A_s$ is the coefficient of the singular term and
$B_s$ the coefficient of the regular one.  It is not difficult to verify
that by leaving the ratio
$$B_s/A_s \equiv \tan \mu_s
\qquad \left( -{\pi \over 2} \leq \mu_s \leq
{\pi \over 2} \right)$$
arbitrary and unspecified one is accommodating all the results of the
self-adjoint extension approach.

One now follows standard scattering theory.  Upon equating the coefficients
of $e^{-ikr}$ terms in (2.4) and a plane wave of the form $e^{-ikr \cos
\phi}$ (i.e., incident from the right), one evaluates $A_s$.  The
scattering amplitude is then the difference between the coefficients of the
asymptotic limits of the $e^{ikr}/r^{1\over 2}$
 term in (2.4) and the plane wave.  One
thus obtains for this amplitude
$$f_j = \left( 2 \pi i k \right)^{-{1 \over 2}} e^{i(\pi - \phi) [N +
{1 \over 2} + {s \over 2}]}
\left\{ e^{2i\delta_s} - 1 \right\} \eqno(2.5)$$
where
$$e^{2i \delta_s} = e^{-i \pi [N+ {1 \over 2} + s/2]}
\ {\exp \{ -{i \pi \over 2}\ [s(\beta - {1 \over 2}) - {1 \over 2}]\}
+ \tan \mu_s \exp \{ {i \pi \over 2}\ [s (\beta -
{1 \over 2}) - {1 \over 2}]\} \over
\exp \{ {i \pi \over 2}\ [s(\beta - {1 \over 2}) - {1 \over 2}]\}
+ \tan \mu_s \exp \{ -{i \pi \over 2}\ [s (\beta -
{1 \over 2}) - {1 \over 2}]\}}\ \ . \eqno(2.6)$$
It is worth emphasizing that one has a two parameter $(\mu_+$ and
$\mu_-$) family of solutions.
Using the step function $\theta (x) \equiv {1 \over 2}\ (1 +
{x \over |x|})$ contact
   with results of ref. 2 is made by
taking $\mu_s = {\pi \over 2} \ \theta (\alpha s)$ which is seen to imply
for (2.5) the form
$$e^{2i\delta_s} = e^{i \pi |\alpha|}$$
which is (remarkably) independent of the
spin parameter $s$.  This means that insofar as spin is concerned the
entire amplitude is described by the factor $\exp [i (\pi - \phi)s/2]$.
This is in fact the matrix appropriate to a rotation by $\pi - \phi$ (i.e.,
the scattering angle) which is necessary to yield a solution consistent
with helicity conservation.  As yet unresolved is the question as to
whether there are other values of $\mu_s$ which are consistent with this
conservation law.  It is to this issue that attention is now directed.

\medskip

\noindent {\bf III. The Helicity Constraint}

\medskip

In order to determine the constraints placed upon the amplitude (2.5) when
helicity conservation is required it is useful to consider an idealized
experiment.  For the sake of simplicity it is prescribed by requiring that
the incoming beam be filtered in such a way as to leave only the
 orbital  angular
momentum  for which $|m + \alpha| < 1$.  The incoming beam is
assumed to be totally polarized along the direction of the unit vector
${\bf n}$ in the plane and the detector is set up to accept
only events along a second direction ${\bf n}^\prime$, also in the
scattering plane.

A convenient tool for this purpose is the projection operator
$$P_{\bf n} = {1 \over 2}\ (1+ \rho \cdot n)$$
where $\rho_1$ and $\rho_2$ are the Pauli matrices which act in the spin
space of the system.  This allows one to write for the cross section
$$\sigma = \ {\rm Tr}\ P_{{\bf n}^\prime} f P_{\bf n} f^* \quad . \eqno(3.1)$$
Now helicity conservation must imply that all scattering events
 which arrive at the
detector will be counted provided that ${\bf n}^\prime$ is a vector which
is rotated by an angle of $\pi - \phi$ (i.e., the scattering angle)
relative to ${\bf n}$.  On the other hand the factor
$$\exp [i (\pi - \phi) )
\rho_3 /2]$$
which occurs in the scattering amplitude is easily seen to have only the
effect of \lq\lq rotating back" this same vector.  Thus (3.1) becomes for
this choice
$$2 \pi k \sigma =\ {\rm Tr}\ P_{\bf n} \left[ e^{2i\delta_s} - 1\right]
P_{\bf n} \left[ e^{-2i \delta_s} - 1 \right] \quad . \eqno(3.2)$$
Even more useful now is to invoke a detector which accepts only helicity
violating (i.e., spin flip) events.  This has the effect of replacing one
of the $P_{\bf n}$'s in (3.2) by $P_{-{\bf n}}$.  One completes the exercise
by setting
$$e^{2i \delta_s} - 1 = a + b \rho_3$$
where $a$ and $b$ are to be determined from Eq. (2.6) and by requiring that
no helicity violating events occur.  One finds that
$$2 \pi k \sigma = |b |^2$$
for this process which thus imposes the requirement that the matrix
$e^{2i \delta_s}$ is proportional to the unit matrix.  It is worth noting
that the choice $\mu_+ = \mu_-$ is not compatible with this condition.  If
it were then one would have the possibility of having regular solutions for
both spin projections and thus at least one choice that would satisfy the
anyon symmetry $\alpha \rightarrow \alpha +1$.  The only solution for
arbitrary flux is $\mu_+ = \mu_- + \pi /2$.  Upon adopting this condition
and setting $\mu_+ = \mu$ Eq. (2.5) becomes
$$f_j = (2 \pi ik)^{- {1 \over 2}} e^{i(\pi -\phi)[N + {1 \over 2} + s/2]}
\left[ e^{2i\delta}-1\right]$$
where
$$\tan \delta = {-1 + (-1)^{N+1} \tan \mu \over 1 + (-1)^{N+1}
\tan \mu} \tan {\alpha \pi \over 2} \quad .$$

The choice $\mu = {\pi \over 2} \theta (\alpha)$
 corresponds to the case considered
in ref. 2.  Since it was based on a physical limiting process one calls it
the physical scattering amplitude $f_p$.  Upon summing over all partial
waves it is seen to have the form
$$f_p = \left( {i \over 2\pi k}\right)^{1/2} {\sin \pi |\alpha | \over \cos
\phi/2}\ e^{i (\pi -\phi) [N + {1 \over 2} + s/2]}
e^{i \phi \epsilon (\alpha)/2} \quad .$$
where $\epsilon (\alpha)$ is the alternating function.  The opposite choice
$\mu = {\pi \over 2} \theta (-\alpha)$ corresponds to the \lq\lq antiphysical"
scattering amplitude
$$f_{\rlap \slash{\rm p}} = - \left( {i \over 2 \pi k}\right)^{{1 \over 2}}
 \ {\sin \pi |\alpha | \over \cos \phi/2} \ e^{i(\pi - \phi)[N +
{1 \over 2} + s/2]} e^{-i \phi \epsilon (\alpha)/2}
\quad .$$
It is \lq\lq antiphysical" in the sense that a repulsive delta function
interaction corresponds to a singular wave function while an attractive one
implies a regular solution.  It is of some interest to observe, however,
that $f_p$ and $f_{\rlap \slash{\rm p}}$ imply that all experiments (even
those using polarized beams) cannot distinguish between these two
amplitudes.  This could only be done if an interference  using a non-AB
interaction term could be made sensitive to the $e^{\pm i\phi \epsilon
(\alpha)/2}$ factor.  Finally, it is to be noted that in the most general
case the total scattering amplitude can also readily
be obtained with the result
$$f = -( 2 \pi ki)^{- {1 \over 2}} \left[ {\cos (\phi/2 - \pi \alpha) \over
\cos \phi /2} - e^{2 i \delta} \right]
e^{i(\pi - \phi)[ N + {1 \over 2}
+ s/2]}$$
where $\delta$ is given by (3.3).

\medskip

\noindent {\bf IV. Conclusion}

\medskip

It is well to discuss with some care the precise sense in which the
$\alpha \rightarrow \alpha + 1$ anyon symmetry fails in the context of this
study.  One should keep in mind that that symmetry is a consequence of the
fact that the partial wave differential equation depends in the spinless
case only on the combination $(m + \alpha)^2$.  As has been realized,
however, the corresponding spin one-half second order differential
equation has a delta function potential which is essentially a Zeeman
interaction.  Its coefficient is proportional to $\alpha s$, a term which
clearly breaks the $\alpha \rightarrow \alpha + 1$ symmetry.  This leads in
the physical $R$ limiting model of ref. 2 to the existence of a singular
solution when $\alpha s < 0$.  From the work of refs. 1 and 7 one knows, of
course, that if one follows the self-adjoint extension method any single
spin component can (at least mathematically) be required to have only a
regular solution.  In the absence of a helicity conservation principle this
can even be done to both components.  Once helicity conservation is invoked
one is free to constrain only one of the two spin values.

The model of ref. 2 yields an amplitude $f_p$ which had its origin in a
solution which was regular (irregular) for $\alpha s > 0 (\alpha s < 0)$.
However, in the self-adjoint extension approach one can exactly reverse
this situation to obtain $f_{\rlap \slash{\rm p}}$ which came from a wave
function which (oddly) has greater concentration at the origin for a
repulsive Zeeman interaction than for an attractive one.  However, in
neither case is the singular solution avoided.  It is merely shifted at
will from one spin component to another.

It has been shown in some detail in ref. 7 that helicity conservation does not
preclude a one parameter family of extensions.  That result has also
emerged in the current study from a somewhat different perspective.
However the claim of ref. 7 that the symmetry $\alpha \rightarrow \alpha
+1$ could thus be preserved in the self-adjoint extension approach has been
seen to be both mathematically and physically untenable.  (In fact no
precise argument for this conclusion is spelled out in that work).  In view
of the considerable interest in the anyonic interpretation in recent years
is is of some importance that this crucial matter not remain uncorrected.

\medskip

\noindent {\bf Acknowledgment}

\medskip

This work has been supported in part by U.S. Department of Energy Grant
No. DE-FG-02-91ER40685.

\vfill\eject

\noindent {\bf References}

\medskip

\item{1.} Ph. de Sousa Gerbert, Phys. Rev. {\bf D40}, 1346 (1989).

\item{2.} C. R. Hagen, Phys. Rev. Lett. {\bf 64}, 503 (1990).

\item{3.} M. G. Alford, J. March-Russell, and F. Wilczek, Nucl. Phys.
 {\bf B328}, 140 (1989).

\item{4.} F. A. B. Coutinho and J. F. Perez, Phys. Rev. {\bf D48}, 932
(1993).

\item{5.} C. R. Hagen, Phys. Rev. {\bf D48}, 5935 (1993).

\item{6.} C. R. Hagen, Phys. Rev. Lett. {\bf 64}, 2347 (1990).

\item{7.} F. A. B. Coutinho and J. F. Perez, Phys. Rev. {\bf D49}, 2092
(1994).

\end